\documentclass[twocolumn,amssymb, nobibnotes, showpacs, superscriptaddress, aps, prd]{revtex4}
\usepackage{amsmath}
\usepackage{epsfig}

\begin{document}
\title{Origin of nonlinear optical processes in matter-wave superradiance}
\author{L. Deng}
\affiliation{Physics Laboratory, National Institute of Standards \& Technology, Gaithersburg, Maryland 20899}
\author{E.W. Hagley}
\affiliation{Physics Laboratory, National Institute of Standards \& Technology, Gaithersburg, Maryland 20899}
\date{\today}

\begin{abstract}
We study a highly efficient, matter-wave amplification mechanism in a longitudinally-excited, Bose-Einstein condensate and reveal a very large enhancement due to nonlinear gain from a six-matter-optical, wave-mixing process involving four photons. Under suitable conditions this optically-degenerate, four-photon process can be stronger than the usual two-photon inelastic light scattering mechanism, leading to nonlinear growth of the observed matter-wave scattering independent of any enhancement from bosonic stimulation. Our theoretical framework can be extended to encompass even higher-order, nonlinear superradiant processes that result in higher-order momentum transfer.      
\end{abstract}

\maketitle

Matter-wave amplification \cite{inouye,kozuma} employing Bose-Einstein condensates has opened a new realm of research where highly efficient inelastic optical wave scattering is employed as a gain mechanism for injection-seeded matter waves.  Although the amplification effect has been observed in both longitudinally-excited
and transversely-excited condensates, the underlying physics of this intriguing and very important process is not yet fully understood. There also has been no convincing explanation of the origin of higher-order, matter-wave scattering that can be subsequently enhanced by bosonic stimulation.

In this letter we investigate the properties of the lowest-order, {\it nonlinear} optical growth process in electromagnetic wave scattering by ultra-cold media such as condensates. We show that the lowest-order process is six-matter-optical-wave-mixing (four optical waves plus two matter waves) where the propagation of an internally-generated optical field plays an important role. We demonstrate that under strong excitation this highly efficient, nonlinear, optical wave-mixing process can be even more important than the usual two-photon scattering mechanism \cite{lu1}.

We study the case of longitudinal excitation, where a linearly-polarized pump laser propagates along the $-\hat{\bf x}$ direction (along the long axis from right to left in Fig. 1a) \cite{lu1}, and a very weak initial field \cite{lu1} starts at the opposite end of the condensate and propagates along the $+\hat{\bf{x}}$ direction (Fig. 1a, red wavy arrow).  
We investigate, in the small signal limit with semi-classical and perturbation approximations, the coherent propagation dynamics of this weak field. We note that a similar nonlinear, wave-mixing enhancement will also occur for transversely-excited (non-coaxial) condensates.

Consider the first-order {\it coherently} scattered matter wave depicted in Fig. 1a that moves with momentum $2\hbar k_L(-\hat{\bf x})$.
Theoretically, contributions to the growth
of this first-order scattered matter wave must include all odd orders of the pump field. In general, odd- (even-) order, matter-wave scattered components involve all odd (even) powers of the pump field. Conceptually, the general expression for the $2\hbar k_L$ matter-wave component is 
\begin{equation}
\sum_{m=0}^{\infty}C_m(a_P)^m(e_P)^m(a_P)(e_G), \nonumber
\end{equation}
where the $C_m$'s are probability amplitudes, $a_P$ ($e_P$) represents absorption (emission) of photons at the pump frequency, respectively, and $e_G$ represents emission into the internally-generated field \cite{ref1}. These internally-generated fields are called end-firing modes in transverse-excitation geometries \cite{inouye}, and backward-propagating modes in longitudinal-excitation geometries \cite{lu1}. For a given $m$ it is clear that the corresponding term represents a $2(m+1)$ photon process that only imparts a net momentum of $2\hbar k_L$ to 
atoms recoiling in the direction of the pump field. As we show below, the gain associated
with the $m=+1$ order process can contribute significantly to the nonlinear growth of the scattered matter wave depicted in Fig 1a.  

\begin{figure}
\centering
\includegraphics[angle=90,width=3in]{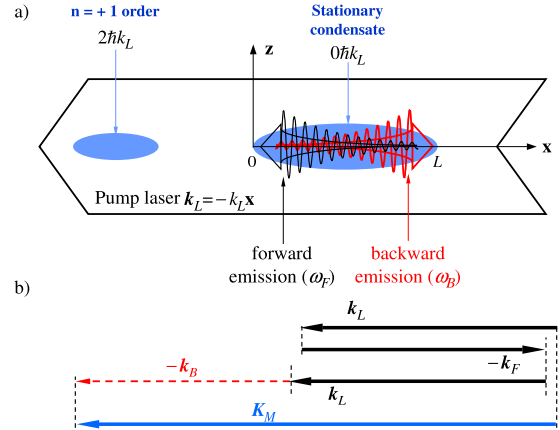} 
\caption{(a) Schematic drawing of lowest-order superradiant scattering with a longitudinally-excited condensate. (b) Multi-matter-optical, wave-mixing vector diagram of the optically-degenerate, four-photon process resulting in $\vec{K}_M=\vec{k}_L-\vec{k}_F+\vec{k}_L-\vec{k}_B$, where $\vec{K}_M$ is the matter wavevector.}
\end{figure}
  
Note that in small-signal theory the leading contribution of all possible higher-order photon processes is the optically-degenerate process depicted in Fig. 1b. Here a pump photon is absorbed and subsequently re-emitted via. stimulated emission into the pump, resulting in negligible momentum transfer to the atom. This is {\it coherently} \cite{shen} followed by the absorption of another pump photon and subsequent stimulated emission into the internally-generated field that counter-propagates the pump field. We stress this is a coherent four-photon process where no population is transfered to the one-, two-, or three-photon intermediate momentum states (Fig. 2). This coherent four-photon cycle is an optically-degenerate, multi-matter-optical, wave-mixing process where atoms acquire only $2\hbar k_{L}$ net momentum and co-propagate with the pump field.  Consequently, this nonlinear process contributes exclusively to the first-order, collectively-scattered matter-wave mode, resulting in nonlinear growth of this mode as the pump intensity increases.  

\begin{figure}
\centering
\includegraphics[angle=90,width=3in]{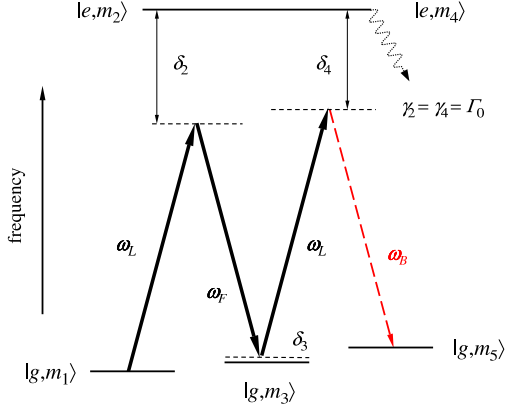} 
\caption{Schematic diagram of energy levels and laser couplings. $|g,m_j\rangle$ and $|e,m_j\rangle$ represent electronic ground
and excited states with momentum states $m_j$, where $j=1,3,\text{and}\,5$ correspond to $0\hbar k_L$, $\approx0\hbar k_L$, and $2\hbar k_L$, respectively.  $\omega_F$ is the frequency of the forward-emitted field stimulated by the pump with $\omega_F\approx\omega_L$. $\omega_B$ is the frequency of the backward-propagating (against the pump) field generated
by the wave mixing process. The one-photon detuning is $\delta=\delta_2=\delta_4$, and the spontaneous emission linewidth of the excited electronic state
is $\gamma_2=\gamma_4=\Gamma_0$.}  
\end{figure}

Two key elements of this four-photon process must be emphasized: (1) the pump-laser-stimulated, forward-emitted photon propagates co-linearly with, and cannot be distinguished from, the pump field; and (2) the pump field (${\bf E}_L$) and the forward-emitted field (${\bf E}_F\approx{\bf E}_L$, $\vec{k}_F\approx\vec{k}_L$, and $\omega_L\approx\omega_F$) form a co-linear-beam, $\Lambda$-type, two-photon excitation that leads to an analytical solution with significant intermediate two-photon enhancement for the internally-generated, backward-propagating weak field (${\bf E}_B$, $\vec{k}_B\approx-\vec{k}_L$).  This results in a three-field, strong-pumping scheme with a fully populated two-photon ``intermediate" electronic state and a very small intermediate two-photon detuning \cite{shen}, both of which significantly enhance the four-photon process.

We start with the Schr$\ddot{o}$dinger equation for the total system wave vector $|\Psi_T\rangle=|\phi\rangle|\psi\rangle$, where $|\phi\rangle$ and $|\psi\rangle$ are the electronic and mean-field macroscopic atomic center-of-motion state vectors, respectively,

\begin{equation}
i\hbar\frac{\partial|\Psi_{T}\rangle}{\partial t}=-\frac{\hbar^2}{2M}\frac{\partial^2|\Psi_{T}\rangle}{\partial {\bf x}^2}+\hat{V}|\Psi_{T}\rangle.
\end{equation}
Here $M$ is the atomic mass, and the four-photon matter-optical field interaction operator is

\begin{eqnarray}
\hat{V}&=&
-\frac{1}{2}\frac{(\hat{{\bf d}}\cdot{\bf E}^{(-)})(\hat{{\bf d}}\cdot{\bf E}^{(+)})(\hat{{\bf d}}\cdot{\bf E}^{(-)})(\hat{{\bf d}}\cdot{\bf E}^{(+)})}{\hbar\Delta_2\hbar\Delta_3\hbar\Delta_4}\nonumber\\
&+&h.c.\quad
\end{eqnarray}
The total electric field amplitude is ${\bf{E}}^{(+)}={\bf{E}}_L^{(+)}e^{-ik_Lx-i(\omega_L-k_L\cdot{v_D})t}+{\bf{E}}_F^{(+)}e^{-ik_Fx-i(\omega_F-k_F\cdot{v_D})t}+{\bf{E}}_B^{(+)}e^{ik_Bx-i(\omega_B+k_B\cdot{v_D})t}$, and the one-photon Doppler recoil velocity is
$v_{D}=\hbar k_{L}/M$.  In addition, ${\bf{E}}_F={\bf{E}}_L$, $\vec{k}_F=\vec{k}_L$, and $\omega_F=\omega_L$ are understood.  We have defined the complex detunings $\Delta_j=\delta_j+i\gamma_j/2$ with $j=2,3,4$.  Later in our calculation we assume the one- and three-photon detunings are the same ($\delta_2=\delta_4=\delta$), and also set $\gamma_2=\gamma_4=\Gamma_0$ ($\Gamma_0$ is the spontaneous emission rate of the upper electronic state), therefore $\Delta_2=\Delta_4=\Delta$. Figure 2 shows the relevant detunings and states in this four-photon scattering process. The Schr$\ddot{o}$dinger equation can now be expressed as
\begin{eqnarray}
\frac{\partial|\psi\rangle}{\partial t}&=&\frac{i\hbar}{2M}\frac{\partial^2|\psi\rangle}{\partial {\bf x}^2}
-iB
|\psi\rangle\nonumber\\
&-&i2A\left[\frac{E_{B}^{(+)}}{E_{L}^{(+)}}+\frac{E_{B}^{(-)}}{E_{L}^{(-)}}\right]|\psi\rangle
-i\omega_{MF}|\psi|^2|\psi\rangle,
\end{eqnarray}
where \begin{eqnarray*}
A\approx-g_0^2\frac{\delta^2}{\delta_3},\quad 
B\approx-ig_0^2\frac{\delta\Gamma_0}{\delta_3},\quad g_0=\frac{|\Omega_L|^2}{|\Delta|^2},
\end{eqnarray*}
$\Omega_L=|d|E_L/\hbar$, $d$ is the transition matrix element, and $\omega_{MF}$ describes the mean-field interaction.
 
Note that the two-photon detuning formed by the pump and the forward-propagating field stimulated by the pump is very small \cite{stenger}. This small two-photon detuning leads to significant enhancement of the coherent four-photon process, making it an important matter-wave scattering mechanism when the pump field is sufficiently intense.  From a nonlinear optics viewpoint this is analogous to four-wave mixing with a nearly resonant, two-photon intermediate state, and it is well known that an intermediate state can significantly enhance the nonlinear interaction strength \cite{shen}.

We expand the collective atomic recoil mode of the macroscopic mean-field atomic wave function as

\begin{eqnarray}
|\psi\rangle&=&\sum_{m,q}\Psi_{m,q}(t)\,e^{-i\omega_qt-im(k_L-k_F)x-iq(k_L+k_B)x}|q\rangle,\quad
\end{eqnarray}
where $\Psi_{m,q}$ represents the amplitude of the $q$-th order collective atomic recoil mode due to a $2(m+1)$-photon optical process. In addition, $\omega_q=(2q)^2n^2\omega_R+\omega_{MF}$ \cite{campbell}, where $\omega_R=\hbar k_L^2/2M$ is the one-photon recoil frequency, and $n$ is the optical refractive index. The four-photon process we are analyzing occurs when $m=q=1$ (the $m=0$, $q=1$ term represents the usual two-photon process). Higher-order optical processes that also transfer $2\hbar k_L$ ($q=1$) of momentum to the scattered condensate occur when $m\geq2$. Note that for first-order, matter-wave scattering $\omega_{q=1}$ is exactly the recoil frequency measured in a recent experiment \cite{campbell} using an interferometric technique \cite{hagley}.

Substituting Eq. (4) into Eq. (3), the equation of motion in the interaction picture becomes

\begin{eqnarray}
\frac{\partial\psi_q}{\partial t}&=&-\left(g_0^2\frac{\delta\Gamma_0}{\delta_3}+\gamma_B+i\omega_{MF}\sum_{q^{'},q^{''}}\psi_{q^{'}}\psi_{q^{''}}^{*}\right)\psi_q\nonumber\\
&+&i2g_0^2\frac{\delta^2}{\delta_3}\frac{E_{B}^{(+)}}{E_{L}^{(+)}}
\psi_{q+1}\,e^{i(2q+1)4n^2\omega_Rt+i\Delta_Lt}\nonumber\\
&+&i2g_0^2\frac{\delta^2}{\delta_3}\frac{E_{B}^{(-)}}{E_{L}^{(-)}}
\psi_{q-1}\,e^{i(2q-1)4n^2\omega_Rt-i\Delta_Lt}.
\end{eqnarray}
In deriving Eq. (5) we focused on the four-photon process ($m=1$) in which absorption of a pump photon and subsequent stimulated emission into the pump field transfers no net momentum ($k_L-k_F\approx0$). Correspondingly, the amplitude of 
the four-photon process of interest is $\psi_{q=1}$.  In Eq. (5) we added a phenomenological decay term  $-\gamma_B\psi_q$ 
to account for the Bragg resonance linewidth in the last step of the four-photon process when the backward field is emitted, 
and $\Delta_L=\omega_L-\omega_B$. 

In our semi-classical perturbation calculation we focus on first-order collective atomic recoil motion under low-intensity, long-pulse excitation.  Therefore, only terms with $q=0,1$ will be included, and the initial condensate density will be treated as undepleted ($|\psi_0|^2\approx n_0= constant$) \cite{luref3}. In this case, to first order, the mean field term is given by $i\omega_{MF}|\psi_0|^2\psi_1$.

The appropriate polarization operator describing the generation of the backward-propagating optical field $E_B$ in the four-photon, optically-degenerate, wave-mixing process is

\begin{eqnarray}
{\bf P}&=&\hat{d}\frac{(\hat{d}\cdot{\bf E})(\hat{d}\cdot{\bf E})(\hat{d}\cdot{\bf E})}{\hbar\Delta_2\hbar\Delta_3\hbar\Delta_4}|\psi|^2.
\end{eqnarray}
Using Eq. (6), the Maxwell equation for $E_B$ can be expressed as

\begin{eqnarray}
\frac{\partial E_{B}^{(+)}}{\partial x}+\frac{1}{c}\frac{\partial E_{B}^{(+)}}{\partial t}&=&-i\frac{\kappa_0}{\Delta}E_{B}^{(+)}\nonumber\\
&+&i\frac{\kappa_{0}}{\Delta}E_{L}^{(+)}P_{+1}\left(g_0\frac{\Delta_2^*}{\Delta_3}\right),
\end{eqnarray}
where $P_{+1}=\psi_{0}\psi_{1}^{*}\,e^{i(4n^2\omega_{R}-\Delta_{L})t}$, and $\kappa_0=2\pi|d|^2\omega_B/(c\hbar)$.

Equations (5) and (7) can be solved analytically using the time Fourier transform method \cite{lu1}, yielding
\begin{eqnarray}
\frac{\partial\epsilon^{(+)}}{\partial x}&=&-i\frac{\kappa_0}{\Delta}n_0\epsilon^{(+)}+i\frac{\omega}{c}\epsilon^{(+)}\nonumber\\
&+&\kappa_0g_0n_0\left(2g_0^2\frac{\delta^2}{\delta_3^2}\right)\left[\frac{1}{-i\omega-D_{+}}\right]\epsilon^{(+)},
\end{eqnarray}
where $\epsilon^{(+)}$ is the Fourier transform of $E_B^{(+)}$, $\omega$ is the transform variable, and

\begin{equation}
D_{+}=i(4n^2\omega_{R}+\omega_{MF}-\Delta_{L})-\left(\frac{g_0^2\delta\Gamma_0}{\delta_3}+\gamma_B\right).
\end{equation}
The first term in Eq. (9) describes the Bragg resonance condition for the two-photon transition involving one pump and one backward-emitted photon with the first-order atomic mean-field shift included.  Thus, the four-photon process involving three pump photons and one backward-emitted photon
generates a collective atomic recoil mode that is identical to the two-photon process.

The possibility of significant enhancement to the backward generated field can be seen by comparing the quantity in parentheses in Eq. (8) with the result for the corresponding two-photon scattering case \cite{lu1}. Clearly, under the Bragg resonance $4n^2\omega_{R}+\omega_{MF}=\Delta_{L}$, and two conditions are required for significant coherent enhancement:

\begin{eqnarray}
\gamma_B\gg\left|\frac{g_0^2\delta\Gamma_0}{\delta_3}\right|\;\text{and}\;\;
2g_0^2\frac{\delta^2}{\delta_3^2}\approx 1.\nonumber
\end{eqnarray}
The first condition implies that the dominate four-photon resonance linewidth is that of the counter-propagating-beam, Bragg-resonance linewidth involving the backward-propagating field $E_B$. The second condition pertains to the enhancement over the usual two-photon process.  We stress that this process, and other higher-order optical processes \cite{ref1}, result in the generation and nonlinear growth characteristics of the scattered matter-wave components under strong pumping. These higher-order, matter-wave seeds, produced by the underlying nonlinear optical processes, are then subsequently amplified by bosonic stimulation.

The gain of the backward-propagating electromagnetic wave in this coherent process can be expressed as

\begin{eqnarray*}
G^{(4)}= G\left(2g_0^2\frac{\delta^2}{\delta_3^2}\right)\left[\frac{g_0\Gamma_0+\gamma_B}{g_0\Gamma_0\left(g_0\frac{\delta}{\delta_3}\right)+\gamma_B}\right],
\end{eqnarray*}
where $G=\kappa_0n_0g_0/(g_0\Gamma_0+\gamma_B)$ is the usual propagation gain of the two-photon scattering process \cite{lu1}. We note that the second-order correction to the nonlinear gain is dependent on the sign of both the pump laser detuning $\delta$, and the two-photon detuning $\delta_3$. The above two requirements for enhancement can be expressed as

\begin{equation}
\left|2\frac{\delta}{\delta_3}\frac{\gamma_B}{\Gamma_0}\right|\gg 2g_0^2\frac{\delta^2}{\delta_3^2}\approx 1,\nonumber
\end{equation}
which can be easily satisfied experimentally. For the experiment reported in \cite{inouye2}, with 

\begin{equation}
2g_0^2\frac{\delta^2}{\delta_3^2}=1,\nonumber
\end{equation}
the four-photon process would be as important as the two-photon process. With $\Gamma_0/2\pi=10^7$ Hz, $|\delta_3|/2\pi=10^4$ Hz, and $|\delta|/2\pi=1.7\times10^9$ Hz \cite{inouye2}, we obtain $g_0=4.2\times 10^{-6}$. This corresponds to a Rayleigh scattering rate of $R=g_0\Gamma_0/4=66$ Hz, which is slightly higher than the scattering rate for the lowest-order data reported in \cite{inouye2}. In addition, with $\gamma_B/2\pi\approx 10^4$ Hz we find $2\delta\gamma_B/(\delta_3\Gamma_0)=340\gg 1$, and the first condition is also well satisfied.  Therefore under longitudinal excitation, first-order, matter-wave scattering can be significantly enhanced by this coherent, nonlinear, four-photon process. This is similar to the situation in nonlinear optics where it is known that pumping a small segment of nonlinear fiber produces an optical four-wave mixing signal that is stronger than concurrent Rayleigh or Raman scattering \cite{agarwal}.

Finally, we comment on matter-wave amplification efficiency and the momentum
spread of the scattered condensates in the two excitation geometries.
Under transverse excitation the thickness of the condensate
in the direction of the pump laser is much less
than the maximum condensate dimension along which the scattered
photons propagate. Thus, from a bosonic stimulation point of view,
the interaction time of the scattered and stationary condensates is limited primarily by the spatial extent of the short axis, $t_T = l_{short}M/(2\hbar k_L)$.
To make matters worse, the spatial extent of the condensate in the radial dimension will be compressed by an induced optical-dipole potential as the process evolves \cite{luref3} with a red detuned pump, further reducing the interaction time. However, with longitudinal-excitation this interaction time is inherently much longer, $t_L = l_{long}M/(2\hbar k_L)$ since $l_{long}\gg l_{short}$, and this can result in higher gain from bosonic stimulation. We further note that with longitudinal-excitation the momentum spread of the scattered condensate can be reduced by adjusting the pump laser divergence to cancel the spread of the backward-generated field.  This, however, is not possible with transverse-excitation because the pump and generated fields propagate perpendicularly.  In a sense, longitudinal
excitation is analogous to a self-injection-seeded, longitudinally-pumped fiber laser which is known to have high gain and narrow bandwidth. These observations are consistent with the observed efficiency and stability benefits \cite{douglas} of matter-wave amplification employing longitudinal-excitation \cite{kozuma} over the transversely-excited geometry \cite{inouye}.

In conclusion, we have revealed an optically-degenerate, four-photon process that can significantly enhance the gain of first-order, matter-wave superradiant scattering. We have shown that this higher-order, nonlinear process benefits from an intermediate two-photon resonant enhancement that can lead to pronounced nonlinear growth of the scattered, first-order, matter-wave mode. This nonlinear growth occurs independent of bosonic stimulation, which is another very important nonlinear gain mechanism that was not treated in this study. We emphasis that our theoretical framework is capable of treating even higher-order, nonlinear optical growth processes that result in higher-order momentum transfer to the condensate. These higher-order, nonlinear optical processes produce the matter-wave seeds that can be subsequently boosted by bosonic stimulation in condensates, leading to dramatic superradiant scattering with red detunings and high pump intensities. Our work clearly shows that the underlying physics of superradiance is multi-matter-optical wave mixing, where higher-order, nonlinear optical processes play a dominate role in the generation of collective atomic recoil motion.


\begin{thebibliography}{13}
\bibitem{ref1} In general, higher-order scattering terms in the Hamiltonian can be easily incorporated into the theoretical framework of Ref. \cite{lu1}.
\end{thebibliography}
\end{document}